\documentclass[aps, floats, showpacs,nofootinbib,floatfix]{revtex4}%
\usepackage{eurosym}
\usepackage{amsfonts}
\usepackage{amsmath}
\usepackage{subfigure}
\usepackage{amssymb}%
\usepackage{epsfig}
{\catcode`\|=\active \gdef\Braket#1{\left<\mathcode`\|"8000\let|\bravert
{#1}\right>}}
\def\bravert{\egroup\,\vrule\,\bgroup}

\newcommand{\beq}{\begin{eqnarray}}
\newcommand{\eeq}{\end{eqnarray}}
\newcommand{\bw}{\begin{widetext}}
\newcommand{\ew}{\end{widetext}}

\begin{document}

\title{ Non-perturbative momentum dependence of the coupling constant and hadronic models}
\author{A. Courtoy}
\email{Aurore.Courtoy@pv.infn.it}
\affiliation{
INFN-Sezione di Pavia, Via Bassi 6, 27100 Pavia, Italy}
\author{S. Scopetta}
\email{Sergio.Scopetta@pg.infn.it}
\affiliation{Dipartimento di Fisica, Universit\`a degli Studi di Perugia, and
INFN, sezione di Perugia, via A. Pascoli
06100 Perugia, Italy}
\author{V. Vento}
\email{Vicente.Vento@uv.es}
\affiliation{
Departamento de F\'{\i}sica Te\'orica and Instituto de F\'{\i}sica Corpuscular,
Universidad de Valencia-CSIC, E-46100 Burjassot (Valencia), Spain.}

\date{\today }

%%%%%%%%%%%%%%%%%%%%%%%%%%%%%%%%%%
\begin{abstract}
Models of hadron structure are associated with a  hadronic scale which allows by perturbative evolution to calculate observables in the deep inelastic region.  
The resolution of Dyson-Schwinger equations leads  to  the freezing of the QCD running coupling (effective charge) in the infrared, which is best understood as a dynamical 
generation of a gluon mass function, giving rise to a momentum dependence which is free from infrared divergences.  We use  this new development to understand why perturbative treatments 
are working reasonably well despite the smallness of the hadronic scale.
\end{abstract}
%%%%%%%%%%%%%%%%%%%%%%%%%%%%%%%%%%

\maketitle

\section{ Introduction}
The constituent quark, one of the most fruitful concepts in 20th century physics, was proposed to explain the structure of the large number of hadrons being discovered
in the sixties . This concept led to the formulation of quark models designed to describe the static hadronic properties \cite{Lipkin:1973nt}. On the other hand, deep inelastic 
scattering (DIS) of leptons off protons was explained in terms of pointlike constituents named partons \cite{Kogut:1972di}. The connection between these  descriptions should allow 
to understand hadronic phenomena at different scales.  To do so a   formalism was developed by Traini et al. \cite{Traini:1997jz} to make models of hadron structure
 predictive in the deep inelastic scattering (DIS) region which  consists of three basic ingredients: a low energy hadronic scale ($\mu_0$), defined by the second moment 
 of the  partonic distribution; the matrix elements of low twist operators defining the observables under scrutiny, which are calculated using  models of hadron structure;
 the evolution of these matrix elements  via the renormalization group equations from the hadronic scale to the high energy scale of DIS data. If the low energy model 
 consists of only valence quarks, the hadronic scale  determined by the above procedure turns out to be rather low ( $\mu_0^2 \sim 0.1$ GeV$^2$) and therefore  
 Next to Leading Order ($NLO$) evolution was applied.

It is well established by now that the QCD running coupling (effective charge)  freezes in the deep infrared. 
This property can be best understood from the point of view of the dynamical gluon mass generation \cite{Cornwall:1982zr,Aguilar:2006gr,Binosi:2009qm}. 
 Even though  the  gluon is massless  at the  level  of the  fundamental  QCD Lagrangian, and  remains massless to all order in perturbation theory, the non-perturbative QCD
dynamics  generate  an  effective,  momentum-dependent  mass,  without affecting    the   local    $SU(3)_c$   invariance,    which   remains
intact. The gluon mass generation is a purely non-perturbative effect \cite{Cornwall:1982zr, Aguilar:2008xm}.

The aim of the present investigation is to justify the  perturbative evolution  approach by comparing  it to the non-perturbative momentum dependence, which we will name for simplicity non- perturbative evolution, as determined by the phenomenon of the freezing of the coupling constant, and to analyze the consequences of introducing an  effective gluon mass.  In section 2, we shall extend the perturbative evolution of the coupling constant to Next-to-Next-to-Leading-Order,  solving it exactly. We will analyze convergence of the perturbative expansion in the low energy region and compare the results to non-perturbative evolution.  In section 3 we will rediscuss the hadronic scale scenario within non-perturbative evolution and in section 4 revisit previous calculations of the Sivers and Boer-Mulders functions at the light of this discussion. We end up by drawing some conclusions.

%%%%%%%%%%%%%%%%%%%%%%%%%%%%%%%%%%
\section{Perturbative evolution versus Non-Perturbative evolution}

Let us describe forst the perturbative evolution of the coupling constant. At $N^mLO$ the scale dependence is given by

$$\frac{d \, a (Q^2)}{ d(\ln \;Q^2)}  = \beta_{N^mLO}(\alpha) =\stackrel{m}{\sum_{k=0}} a^{k+2} \beta_k,$$
where 
$$a = \frac{\alpha}{4 \pi}.$$
The expansion coefficients of the $\beta$-function are known up to $k=3$ \cite{Larin:1993tp}, however we will only show here the solution to $k=2$, i.e., $N^2LO$
\begin{eqnarray}
\beta_0 & = &\;\; 11\;\; - \;\;\;\frac{2}{3}\, n_f \nonumber \\
\beta_1 & = & \;102\;\, -\; \;\frac{38}{3} \,n_f \nonumber \\
\beta_2 & = & \frac{2857}{2}- \frac{5033}{18}  n_f  +  \frac{325}{54}  n_f^2, 
\end{eqnarray}
where $n_f$ stands for the number of effectively massless quark flavors and $\beta_k$ denote the coefficients of the usual four-dimensional $\overline{MS}$ beta function of QCD.
The evolution equations for the coupling constant can be integrated out exactly leading to\\
%%%%%%%%%%%%%%%%%%%%%%%%%%%%%%%%%%%%%%%%%%%%%%%%%%%%%%%%
%          Fig.1 running of the  effective coupling perturbatively
%%%%%%%%%%%%%%%%%%%%%%%%%%%%%%%%%%%%%%%%%%%%%%%%%%%%%%%%
\begin{center}
\begin{figure}[b]
\includegraphics[scale= 1.2]{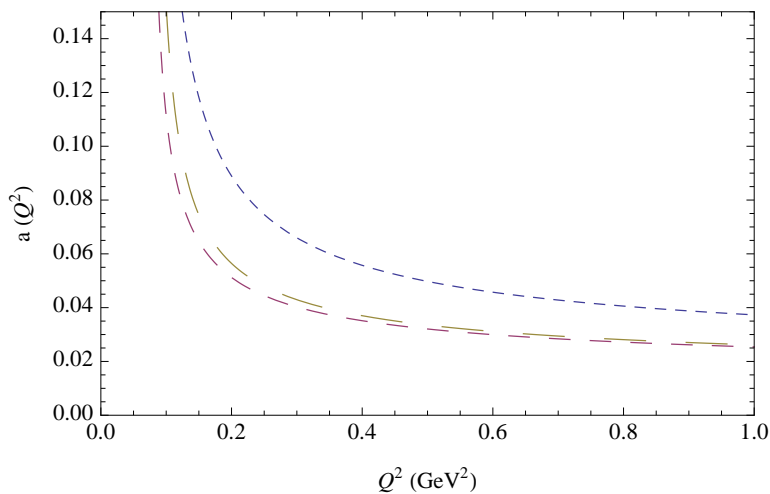}
\caption{ The running of the  coupling constant. The short dashed curve curve corresponds to the Leading Order ($LO$) solution, the medium dashed curve to Next to Leading Order ($NLO$) solution and the long dashed curve to the Next-to-Next-Leading Order ($NNLO$) solution. We have used in this plot the same value of $\Lambda = 250$ MeV. Note that below $0.1$  GeV$^2$ the curves  approach their Landau pole.}
\label{aperb} 
\end{figure}
\end{center}
%%%%%%%%%%%%%%%%%%%%%%%%%%%%%%%%%%%%%%%%%%%%%%%%%%%%%%%%%%%%%%%

\hspace{2cm}\begin{minipage}{13cm}
$ \ln \; (Q^2/\Lambda_{LO}^2)\; \;\;\;\;=\; \;\frac{1}{\beta_0 a_{LO}}$ 

$ \ln (Q^2/\Lambda_{NLO}^2) \;\;\;\, = \; \frac{1}{\beta_0 a_{NLO}} +  \frac{b _1}{\beta_0} \ln (\beta_0 a_{NLO})- \frac{b _1}{\beta_0} \ln (1 + b_1 a_{NLO}) $

$ \ln \; (Q^2/\Lambda_{NNLO}^2) = \frac{1}{\beta_0 a_{NNLO}} +  \frac{b _1}{\beta_0} (\ln  (\beta_0 a_{NNLO})  - \frac{b_1}{2 \beta_0} \ln\;(1 + b_1 a_{NNLO} + b_2 a_{NNLO}^2) \; + $\\
 
\hspace{2.8cm}$\frac{2 b_2  - b_1^2}{2 \beta_0^2 b_2}  \arctan{\frac{2 b_2 a_{NNLO} + b_1}{\sqrt{4 b_2  - b_1^2}} } , $         
\begin{equation}
\label{exact}
 \end{equation}
 \end{minipage}\\[0.3cm]
where $b_k = \frac{\beta_k}{\beta_0}$. These equations, except the first, do not admit closed form solution for the coupling constant, and we have solved them numerically. We show their solution, for the same value of $\Lambda = 250$ MeV, in  Fig. \ref{aperb}.

%%%%%%%%%%%%%%%%%%%%%%%%%%%%%%%%%%%%%%%%%%%%%%%%%%%%%%%%
%          Fig.2 running of the  effective coupling perturbatively
%%%%%%%%%%%%%%%%%%%%%%%%%%%%%%%%%%%%%%%%%%%%%%%%%%%%%%%%
\begin{center}
\begin{figure}[htb]
\includegraphics[scale= 1.05]{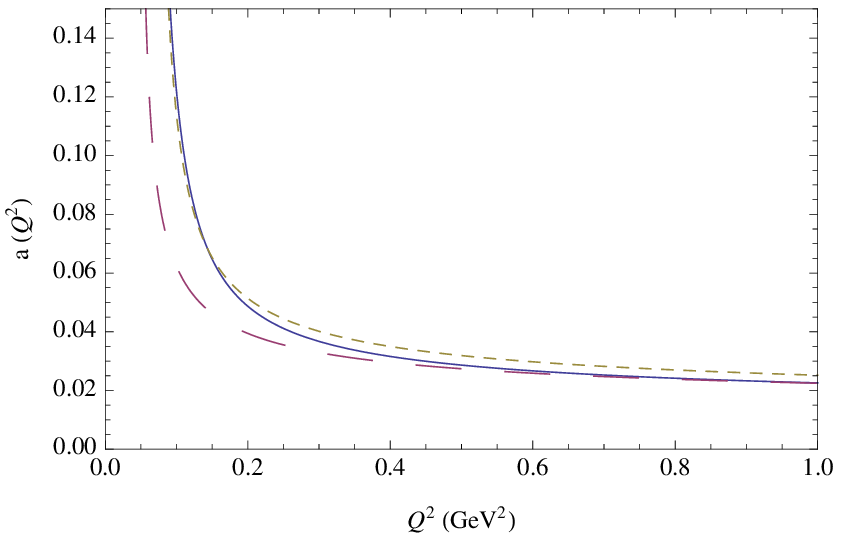}
\includegraphics[scale= 1.05]{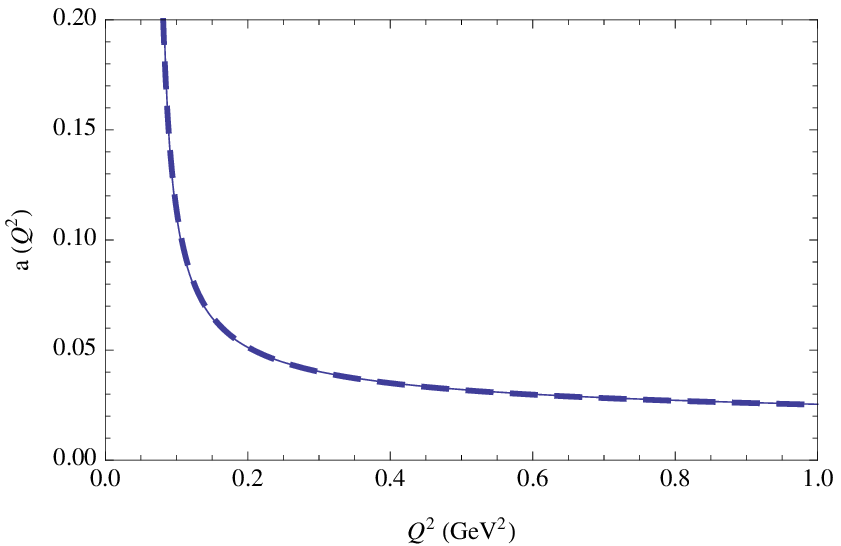}
\caption{The running of the coupling. Left: The solid curve represents  the commonly used approximate iterative solution with $\Lambda= 200$ MeV (see  Eq.(22) of 
ref. \cite{Khorramian:2009xz}, where the NLO approximation is given by the first line of the equation). The long dashed curve shows the exact calculation with the same value of $\Lambda=200$ MeV. Finally the short dashed curve represents the exact calculation with $\Lambda = 250$ MeV. Right: The solid curve represents  the $NLO$ solution  with $\Lambda= 250$ MeV, while the  long dashed curve the $NNLO$ solution with a value of $\Lambda = 235$ MeV. }
\label{comparisons}
\end{figure}
\end{center}
%%%%%%%%%%%%%%%%%%%%%%%%%%%%%%%%%%%%%%%%%%%%%%%%%%%%%%%%%%%%%%%

We show in Fig.\ref{comparisons} (left) studies of the running of the coupling constant for two values of $\Lambda$ in order to compare the exact solution with the commonly 
used iterative solution at $NLO$. We notice that for small values of $Q^2$  the two differ considerably as noted also in a different context by ref. \cite{Shaikhatdenov:2009xd}.

We see in Fig.\ref{aperb} that the $NLO$ and $NNLO$ solutions agree quite well even at very low values of $Q^2$ and in  Fig.\ref{comparisons} (right) that they agree  very well if we change  the value of $\Lambda$ for the $NNLO$ slightly, confirming the fast convergence of the expansion.  This analysis concludes, that even close to the Landau pole, the convergence of the perturbative expansion is quite rapid, specially if we use a different value  of $\Lambda$ to describe the different orders, a feature which comes out from the fitting procedures.

\begin{center}
\begin{figure}[h]
\includegraphics[scale= 1.2]{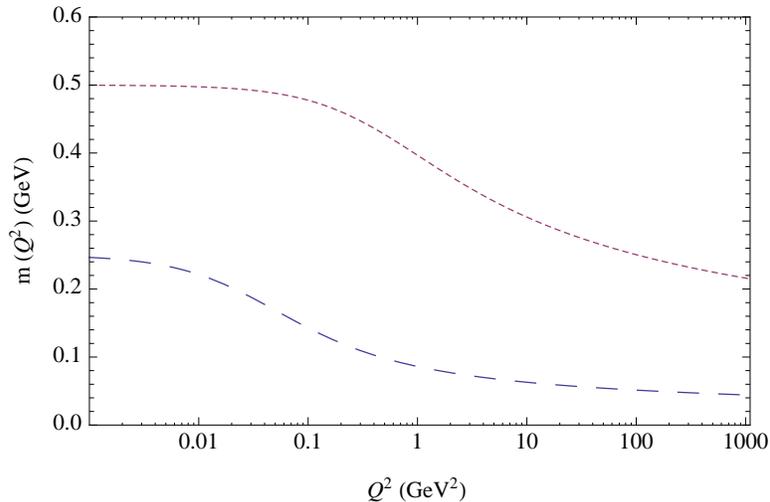}
\caption{ The dynamical gluon mass with a logarithmic running for two sets of parameters, the small mass scenario ($\Lambda =250$ MeV, $m_0 = 250$ MeV, $\rho =1.5$) is shown by the dashed curve; the high mass scenario ($\Lambda =250$ MeV, $m_0 = 500$ MeV, $\rho =2.0$) by the dotted curve.}
\label{fmass}
\end{figure}
\end{center}

It is well established by now that the QCD running coupling (effective charge)  
freezes in the deep infrared. 
This non perturbative property can be best understood from the point of view of
the dynamical gluon mass generation \cite{Cornwall:1982zr,Aguilar:2006gr,Binosi:2009qm,Aguilar:2008xm}. 
 Even though  the  gluon is massless  at the  level  of the  fundamental  QCD Lagrangian, and  remains
massless to all order in perturbation theory, the non-perturbative QCD
dynamics  generate  an  effective,  momentum-dependent  mass,  without
affecting    the   local    $SU(3)_c$   invariance,    which   remains
intact.  
At the level of the Schwinger-Dyson equations
the  generation of such a  mass is associated with 
the existence of 
infrared finite solutions for the gluon propagator, 
i.e. solutions with  $\Delta^{-1}(0) > 0$ .
Such solutions may  
be  fitted  by     ``massive''  propagators  of   the form 
$\Delta^{-1}(Q^2)  =  Q^2  +  m^2(Q^2)$;
$m^2(Q^2)$ is  not ``hard'', but depends non-trivially  on the momentum  transfer $Q^2$.
One physically motivated possibility, which we shall use in here, is  the so called logarithmic mass running, which is defined by

\begin{equation}
m^2 (Q^2)= m^2_0\left[\ln\left(\frac{Q^2 + \rho m_0^2}{\Lambda^2}\right)
\bigg/\ln\left(\frac{\rho m_0^2}{\Lambda^2}\right)\right]^{-1 -\gamma}.
\label{rmass}
\end{equation}
Note that when $Q^2\to 0$ one has $m^2(0)=m^2_0$. Even though in principle we do not have 
any theoretical constraint that would put an upper bound to the value of $m_0$, 
phenomenological estimates place it in the range $m_0 \sim \Lambda - 2 \Lambda$~\cite{Bernard:1981pg,Parisi:1980jy}. The other parameters were fixed at $\rho \sim 1-4$,  ${(\gamma)} = 1/11$ \cite{Cornwall:1982zr, Aguilar:2007ie,Aguilar:2009nf}. The (logarithmic) running of $m^2$, shown in Fig. \ref{fmass} for two sets of parameters, is   associated with the 
the {\it gauge-invariant non-local} condensate of dimension two obtained through the minimization
of $\int d^4 x ( A_{\mu})^2$ over all gauge transformations
\cite{Gubarev:2000eu}.

%%%%%%%%%%%%%%%%%%%%%%%%%%%%%%%%%%%%%%%%%%%%%%%%%%%%%%%%
%          Fig.3 running masses
%%%%%%%%%%%%%%%%%%%%%%%%%%%%%%%%%%%%%%%%%%%%%%%%%%%%%%%%
%%%%%%%%%%%%%%%%%%%%%%%%%%%%%%%%%%%%%%%%%%%%%%%%%%%%%%%%%%%%%%%

%%%%%%%%%%%%%%%%%%%%%%%%%%%%%%%%%%%%%%%%%%%%%%%%%%%%%%%%
%          Fig.4 running of the  effective coupling
%%%%%%%%%%%%%%%%%%%%%%%%%%%%%%%%%%%%%%%%%%%%%%%%%%%%%%%%
\begin{center}
\begin{figure}[b]
\includegraphics[scale= 1.2]{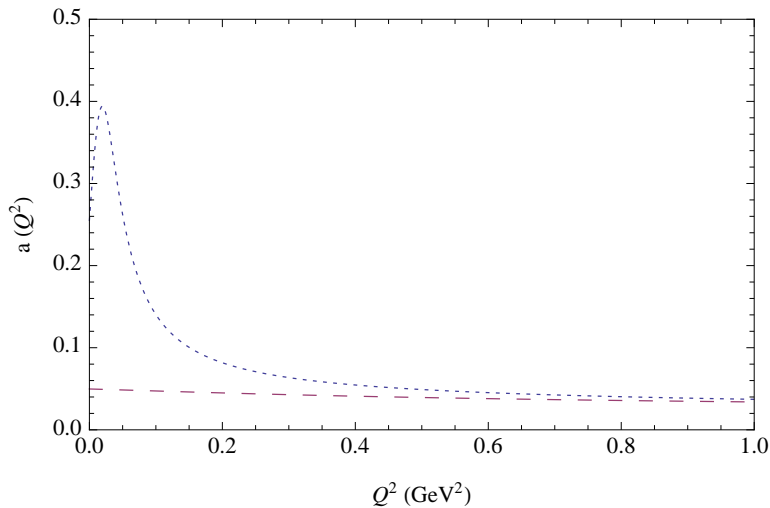}
\caption{ The running of the effective coupling. The dotted curve corresponds to the low mass set of parameters ($m_0= 250$ MeV, $\rho =1.5$, $\Lambda = 250$); the dashed curve to the high mass set of parameters ($m_0= 500$ MeV, $\rho =2.$ and $\Lambda = 250$ MeV). }
\label{alphaJ}
\end{figure}
\end{center}
%%%%%%%%%%%%%%%%%%%%%%%%%%%%%%%%%%%%%%%%%%%%%%%%%%%%%%%%%%%%%%%

The strong coupling constant plays a central role in the evolution of parton densities. The  non-perturbative  generalization  of $\alpha(Q^2)$
the  QCD  running  coupling , comes in the form
\begin{equation}
a_{NP}(Q^2) = \left[\beta_0 \ln \left(\frac{Q^2 +\rho m^2(Q^2)}{\Lambda^2}\right)\right]^{-1} ,
\label{alphalog}
\end{equation}
where we use the same notation as before. $a =\frac{\alpha}{4 \pi}$ and $NP$ stands for Non-Perturbative. Note that its zero gluon mass limit leads to the LO perturbative 
coupling constant momentum dependence.
The $m(Q^2)$ in the argument of the logarithm 
tames  the   Landau pole, and $a(Q^2)$ freezes 
at a  finite value in the IR, namely  
\mbox{$a^{-1}(0)= \beta_0 \ln (\rho m^2(0)/\Lambda^2)$} \cite{Cornwall:1982zr,Aguilar:2006gr,Binosi:2009qm} as can be seen in Fig. \ref{alphaJ} for the same two sets of parameters.

\begin{center}
\begin{figure}[t]
\includegraphics[scale= 1.2]{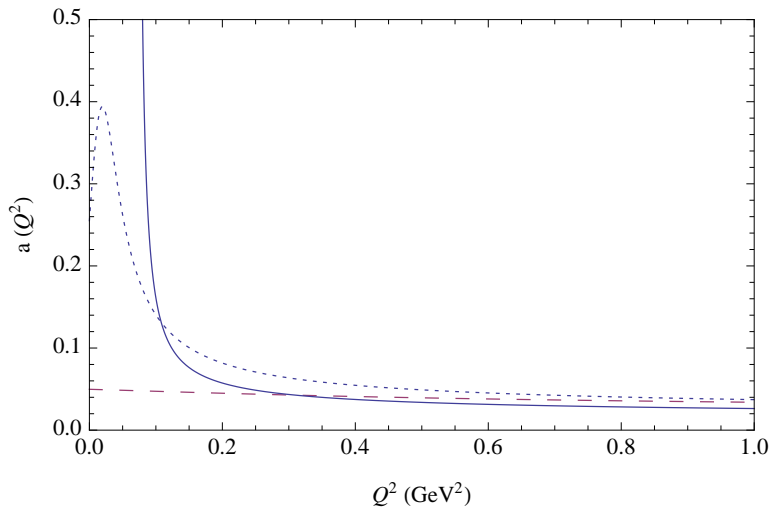}
\caption{ The running of the effective coupling. The dotted and dashed curves represent the non-perturbative evolution with the parameters used above. The solid curve shows the $NNLO$ evolution with $\Lambda = 250$ MeV.}
\label{alpha}
\end{figure}
\end{center}

We discuss in the next section  the relation between the perturbative and non perturbative approaches from the point of view of the hadronic models. Here we note their
numerical similarity.  As shown in Fig.\ref{alpha}, the coupling constant in the perturbative and non-perturbative approaches  are close in size for reasonable values of the parameters from very low $Q^2$ onward ( $Q^2 > 0.1$ GeV$^2$). This result supports the perturbative approach used up to now in model calculations, since it shows, that despite the vicinity of the Landau pole to the hadronic scale, the perturbative expansion is quite convergent and agrees with the non perturbative results for a wide range of parameters.

%%%%%%%%%%%%%%%%%%%%%%%%%%%%%%%%%%%%%%%%%%%%%%%%%%%%%%%%
%          Fig.5 running of the  effective coupling  perturbative and non-perturbative
%%%%%%%%%%%%%%%%%%%%%%%%%%%%%%%%%%%%%%%%%%%%%%%%%%%%%%%%
%%%%%%%%%%%%%%%%%%%%%%%%%%%%%%%%%%%%%%%%%%%%%%%%%%%%%%%%%%%%%%%

%%%%%%%%%%%%%%%%%%%%%%%%%%%%%%%%%%
\section{The hadronic scale}

Models of hadron structure are associated with a momentum scale \cite{Jaffe:1980ti}, the so called hadronic scale \cite{Traini:1997jz}. The hadronic scale is defined at a point where the partonic content of the model, defined through the second moment of the parton distribution,  is known. In the extreme case, i.e., when we assume that the partons are pure valence quarks, the hadronic scale is $\mu_0^2 \sim 0.1$ GeV$^2$.  

Let us see how to understand the hadronic scale in the language of models of hadron structure.  We  use, to clarify the discussion, the original bag model, in its most naive description, consisting of a cavity of perturbative vacuum surrounded by non-perturbative vacuum \cite{Chodos:1974je,Chodos:1974pn}. The bag model has the advantage that it is a field theory  in a cavity and therefore we can use a field theoretic language which is very appealing. The non-perturbative vacuum is endowed with a pressure which keeps the cavity size finite. The quarks and gluons are modes in the cavity satisfying certain boundary conditions. The most simple scenario consists only of valence quarks (antiquarks) as Fock states building a colorless state describing the hadron. The interaction of these Fock states is either with the boundary (confining interaction) or among themselves in the interior via perturbative QCD. Thus even though there might be no gluons in the Fock space, virtual gluons appear through the propagators describing the interaction between the quarks. However these gluons are at least order $\alpha$, while the matrix elements of the quark Fock states can be order $1$. Also one can have quark propagators and therefore one can generate a quark sea via the perturbative QCD interaction \cite{Maxwell:1981kg}. 

The bag model is designed to describe fundamentally static properties, but in QCD all matrix elements must have a scale associated to them as a result of the Renormalization Group (RG)  of the theory.  A fundamental step in the development of the use of hadron models for the description of  properties at high momentum scales was the assertion that all calculations done in a model should have a  RG scale associated to it \cite{Jaffe:1980ti}. Note that the momentum distribution inside the hadron is only related to the hadronic scale and not to the momentum governing the RG equation. Thus a model calculation only gives a boundary condition for the RG evolution as can be seen for example in the $LO$ evolution equation for the moments of the valence quark distribution 

\begin{equation}
<q_v(Q^2)>_n = <q_v(\mu_0^2)>_n \left(\frac{\alpha{(Q^2)}}{\alpha{(\mu_0^2)}}\right)^{d^n_{NS}},
\label{moments}
\end{equation}
where $d^n_{NS}$ are the anomalous dimensions of the Non Singlet distributions.  Inside, the dynamics  described by the model is unaffected by the evolution procedure, and the model provides only the expectation value,  $<q_v(\mu_0^2)>_n$, which is associated with the hadronic scale.  What is the meaning of the hadronic scale?  It is related to the maximum wavelength at which the structure begins to be unveiled, i.e.
$$\lambda \sim \mu_0^{-1} =\sim 0.5 fm.$$
For wavelengths smaller than this one, i.e.  momentum scales higher than $\sim 350$ MeV, the underlying structure unveils. In the most naive scheme, it does in the form of  current quarks, sea quarks and gluons produced solely by evolution. 

This explanation goes over to non-perturbative evolution. The non-perturbative solution of the Dyson Schwinger equations results in the appearence of an infrared cut-off in the form of a gluon mass which determines the finiteness of the coupling constant in the infrared. The mass, interpreted as a Meissner effect \cite{Mandelstam:1980ii}, leads to a confinement mechanism allowing color not  to leak out. A gluon mass interpretation of the MIT bag boundary condition would require an infinite mass, since the color fields do not penetrate  the outside region. Thus the real confinement mechanism is softer than that of the bag, more  line with potential models. In any case, the crucial statement is that the gluon mass does not affect the dynamics inside the bag, where perturbative physics is operative and therefore our gluons inside will behave as massless. However, this mass will affect the evolution as we have seen in the case of  the coupling constant. The generalization of the coupling constant results to the structure function imply that the $LO$ evolution Eq.(\ref{moments})  simply changes by incorporating  the non-perturbative coupling constant  evolution Eq.(\ref{alphalog}). 

%%%%%%%%%%%%%%%%%%%%%%%%%%%%%%%%%%%%%%%%%%%%%%%%%%%%%%%%
%          Fig.6 evolution of the second moment
%%%%%%%%%%%%%%%%%%%%%%%%%%%%%%%%%%%%%%%%%%%%%%%%%%%%%%%%
\begin{center}
\begin{figure}[h]
\includegraphics[scale= 1.2]{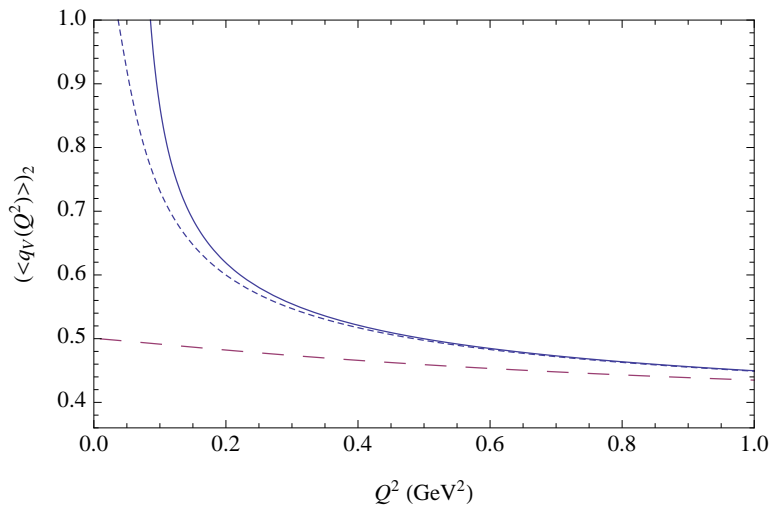}
\caption{ The evolution of the second moment of the valence quark distribution. The solid curve represents the perturbative Leading Order approximation. The dotted  curve and dashed curve represent the  non-perturbative evolution. The parameters are those used  before.}
\label{qval}
\end{figure}
\end{center}
%%%%%%%%%%%%%%%%%%%%%%%%%%%%%%%%%%%%%%%%%%%%%%%%%%%%%%%%%%%%%%%

The non-perturbative results, using the same parameters as before, are quite close to those of the perturbative scheme and therefore we are confident that the latter is a very good approximate description.  We note however, that the corresponding hadronic scale, for the sets of parameters chosen, turns out to be slightly smaller than in the perturbative case ($\mu_0 \sim 0.1$ GeV$^ 2$), even  for small gluon mass $m_0 \sim 250$ and small $\rho \sim 1$.    One could reach a pure valence scenario at higher $Q^2$ by forcing the parameters but at the price of generating a singularity in the coupling constant in the infrared associated with the specific logarithmic form of the parametrization. We feel that this strong parametrization dependence and the singularity are non physical since the fineteness of the coupling constant in the infrared is a wishful outcome of the non-perturbative analysis. In this sense, the non-perturbative approach  seems to favor a scenario where at the hadronic scale we have not only valence quarks but also gluons and sea quarks \cite{Scopetta:1997wk,Scopetta:1998sg}. We mean by this statement that to get a scenario with only valence quarks we are forced to very low gluon masses and very small values $\rho$, while a non trivial scenario allows more freedom in the choice of parameters.

%%%%%%%%%%%%%%%%%%%%%%%%%%%%%%%%%%%%%%%%%%%%%%%%%%%%%%%%%%%%%%%

\section{Nonperturbative Evolution and Final State Interactions.}

Recently the perturbative evolution formalism has been applied to describe the behavior of the T-odd Transverse Momentum Dependent parton distribution functions (TMDs) \cite{Courtoy:2008vi, Courtoy:2008dn,Courtoy:2009pc}. The TMDs are the set of functions that depend on both the Bjorken variable $x$, the intrinsic transverse momentum of the quark $\vec{k}_{\perp}$
and on the scale $Q^2$. The TMDs are fixed by the possible scalar structures allowed by hermiticity, parity and time-reversal invariance. The existence of leading twist final state interactions allows for time-reversal odd functions~\cite{Brodsky:2002cx}. Thus by relaxing time-reversal invariance, one defines two additional functions, namely, the Sivers and the Boer-Mulders function \cite{Sivers:1989cc,Boer:1997nt}. These functions are related, respectively, to single spin and azimuthal asymmetries, and are therefore important in our quest for the understanding of the proton spin.

In the standard approach towards an evaluation of the T-odd distribution functions, the final state interactions are  mimicked  by a one-gluon-exchange. This gluon exchange is usually described through the inclusion of a perturbative gluon propagator\cite{Yuan:2003wk,Courtoy:2008vi}. It is precisely due to this mechanism that these functions have an explicit dependence in the coupling constant and therefore they are ideal to analyze the physical impact of our discussion. Since perturbative QCD governs the dynamics inside the confining region, there is no need to include a non-perturbative massive gluon  in the form given by (\ref{rmass}), inside the bag.  The main effect of the non-perturbative approach consists in a change of the hadronic scale $\mu_0^2$ and the value of the running coupling constant at that scale, as  clearly illustrated in Figs.~\ref{alpha} and \ref{qval}. This leads to a rescaling of the Sivers and BM functions through a change of $\alpha (\mu_0^2)$.

In our previous calculations \cite{Courtoy:2008vi, Courtoy:2008dn, Courtoy:2009pc}, we have used the $NLO$ perturbative evolution, with
\beq
a_s(\mu_0^2) \sim 0.1.
\eeq
Although a solution with this small $a_s$ can be found, with our choice of parameters,  we see, from Figs.~\ref{alpha} and \ref{qval}, that the coupling constant at the hadronic scale in the non-perturbative  approach and in the $NNLO$ evolution  is consistently larger and lies in the interval
\beq
0.1 < a_s(\mu_0^2) < 0.3.
\label{alphalog_mu0}
\eeq

\begin{center}
\begin{figure}[t]
\includegraphics[scale= 0.8]{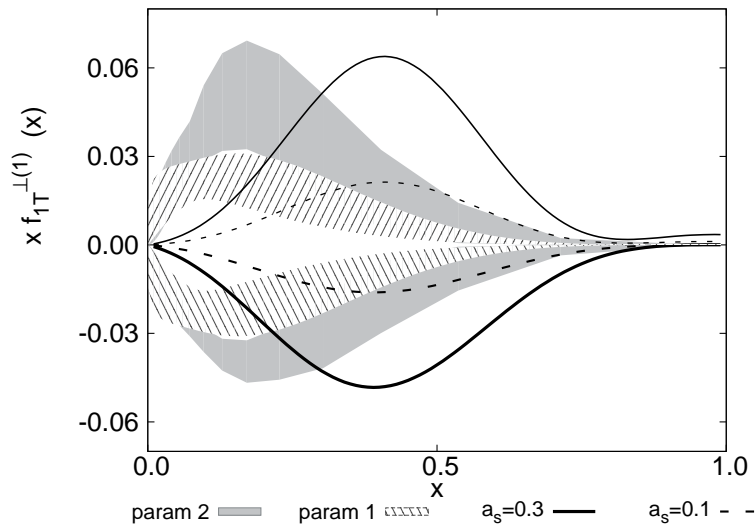}
\caption{The first moment of the Sivers function. The results are given for both the $u$ (thick) and $d$ (normal) distributions. The solid (dashed) curves represent the calculation for $a_s=0.3 (0.1)$. The bands represent the error band for, respectively, the extraction \cite{Collins:2005ie} (full) 
and \cite{Anselmino:2008sga} (stripes).}
\label{sivers}
\end{figure}
\end{center}

Taking into account this range we show the first moments of the Sivers function in  Fig.~\ref{sivers}, where we have two extractions from the data at the SIDIS scale. In order to be able to compare our results to  phenomenology, one should apply the QCD evolution equations. The latter are not available for the TMDs, but the behavior of the known evolutions of the pdfs, lead us to expect that the maximun  of our model results will be shifted by evolution towards lower values of $x$ and the shape of the function will diminish in size, as the experimental extractions seem to indicate.  Moreover, the experiments seem to favor a large  coupling constant, as we have seen, a possibility contemplated by non-perturbative evolution and also by $NNLO$ evolution.

%%%%%%%%%%%%%%%%%%%%%%%%%%%%%%%%%%%%%%%%%%%%%%%%%%%%%%%%
%          Fig.7 sivers
%%%%%%%%%%%%%%%%%%%%%%%%%%%%%%%%%%%%%%%%%%%%%%%%%%%%%%%%
%%%%%%%%%%%%%%%%%%%%%%%%%%%%%%%%%%%%%%%%%%%%%%%%%%%%%%%%%%%%%%%

\begin{center}
\begin{figure}[t]
\includegraphics[scale= 0.8]{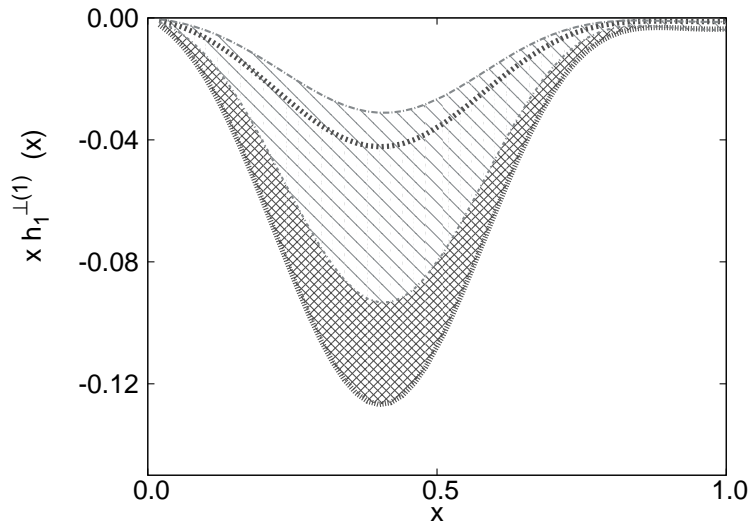}
\caption{The results for the first moment of the Boer-Mulders function calculated within the band $0.1 < a_s < 0.3$. The results are given for both the $u$ (band between the dotted lines) and $d$ (band with  stripes) distributions. Notice that both have the same sign and therefore the bands overlap.}
\label{bm}
\end{figure}
\end{center}

If we apply the same band of values of the coupling constant at the hadronic scale to calculation of the Boer-Mulders function we find the results of Fig. \ref{bm}.  We see thus how the naive scenario may serve to predict new observables and determine their experimental feasibility. 

%%%%%%%%%%%%%%%%%%%%%%%%%%%%%%%%%%%%%%%%%%%%%%%%%%%%%%%%
%          Fig.8 boers-mulders
%%%%%%%%%%%%%%%%%%%%%%%%%%%%%%%%%%%%%%%%%%%%%%%%%%%%%%%%
%%%%%%%%%%%%%%%%%%%%%%%%%%%%%%%%%%%%%%%%%%%%%%%%%%%%%%%%%%%%%%%

The T-odd TMDs have been evaluated in a few models. In most of the models found in the literature  final state interactions are approximated by taking 
into account only the leading contribution due to the one-gluon exchange mechanism \cite{Yuan:2003wk,Courtoy:2008vi, Courtoy:2008dn,Courtoy:2009pc,Pasquini:2010af}. Recently nonperturbative evaluations of the T-odd functions have been proposed, e.g. instanton inclusion~\cite{Cherednikov:2006zn} and nonperturbative eikonal methods ~\cite{Gamberg:2009uk, Gamberg:2010xi}. In the latter references, the authors use a nonperturbative gluon propagator, resulting from a Dyson-Schwinger framework~\cite{Fischer:2006ub}, going therefore  beyond one-gluon-exchange approximation by resumming all order contributions. 
It is worth noticing that the implementation of the final state interactions is model dependent. The discussion we have presented in this paper is not applicable in general to every scheme. The implementation of the nonperturbative evolution as discussed here might be more complex in other (fully nonperturbative) schemes as well as the description of the confinement mechanism.

%%%%%%%%%%%%%%%%%%%%%%%%%%%%%%%%%%%%%%%%%%%%%%%%%%%%%%%%%%%%%%%

\section{Conclusions}

The careful analysis of the previous sections shows that the hadronic scale is close to the infrared divergence (Landau pole) of the coupling constant for conventional values of $\Lambda$.  However, even in this vicinity, the convergence of the $N^mLO$ series is very good for the same $\Lambda$ and small modifications of it provide an extremely precise agreement  for all values of $Q^2$ to the right of the pole. Moreover,  an exciting result is that the values obtained by perturbative QCD with reasonable parameters as defined by DIS data, agree with the non-perturbative evaluation of the coupling constant, which is infrared finite, for  parameters which have been chosen to satisfy lattice QCD restrictions of the propagator, with low values of the gluon mass $m_0 \sim 250$ MeV, and $\rho \sim 2$.

It is interesting to see how the non perturbative framework applies in a simple  way to the evaluation of the T-odd TMDs in the bag model.
This observation confirms the consistency of our previous calculation within this hadronic model. It enables us to control the physics of the problem from 
the model side as well as to infer from the evolution scenarios that, as expected, the naive pure valence quark scenario is not favoured. However, it also shows 
that the naive scenario may well serve to make predictions, within a reasonably small band, which should not be far from experimental expectations.

To conclude we summarize the three main results that we have obtained. The first is that the perturbative approach converges quite well at low momenta despite the fact that the coupling constant is not small. Moreover, since different $N^mLO$ orders carry different values of $\Lambda$,  the convergence is even much better at low momenta than obtained mathematically from the perturbative series. The hadronic scale can be interpreted not only from the point of view of perturbative evolution, but also from that of non-perturbative momentum dependence of the coupling constant and therefore the second result is that the non-perturbative approach provides an explanation of why the evolution from a low hadronic scale, even in the neighbourhood of the Landau pole,  is consistent and can be trusted.  The third result  is that our analysis favors non naive models with a complex quark-gluon structure, however, it also shows that naive models might be used to establish experimental limits of future observations within a small band of values. Moreover,  the evolution makes this band of possible observations much narrower at high momentum,  and thus the predictions become more quantitative.

%%%%%%%%%%%%%%%%%%%%%%%%%%%%%%%%%%
\acknowledgments

We thank A. Aguilar and J. Papavassiliou for illuminating discussions on their work. We also thank Alessandro Bacchetta and Barbara Pasquini for helpful comments. 
This work was carried out with the support of  "Partonic structure of nucleons, mesons and light nuclei" an INFN (Italy) -- MICINN (Spain) exchange agreement. VV was also supported  by HadronPhysics2, a FP7-Integrating Activities and Infrastructure Program of the European Commission under Grant 227431, by the MICINN (Spain) grant FPA2010-21750-C02-01 and by GVPrometeo2009/129.

\end{document}